\documentclass[preprint,showpacs,nofootinbib,prd,aps]{revtex4-1}
\usepackage[utf8]{inputenc}
\usepackage{graphicx,amstext,amssymb}

\begin{document}
\title{Bulk observables in the LHC 5.02 TeV Pb+Pb collisions within the integrated HydroKinetic Model}

\author{V.~M.~Shapoval$^{1}$}
\author{Yu.~M.~Sinyukov$^{1}$}

\affiliation{$^1$Bogolyubov Institute for Theoretical Physics,
03143 Kiev,  Ukraine}

\begin{abstract}
The paper is devoted to the description and prediction of various bulk observables
in the Pb+Pb collisions at the LHC energy $\sqrt{s_{NN}}=5.02$~TeV within the integrated hydrokinetic model (iHKM).  
Sensitivity of the results to the choice of the appropriate model parameter values is also investigated. It is found that 
changing of the relaxation time and the rate of thermalization, which characterize the pre-thermal stage of the matter evolution, 
as well as switching to another equation of state at the hydrodynamic stage and the corresponding hadronization temperature, 
does not destroy the results, if simultaneously one provides an appropriate adjusting of the initial time for the superdense matter 
formation and related maximal initial energy density. 
\end{abstract}

\pacs{13.85.Hd, 25.75.Gz}
\maketitle

Keywords: {\small \textit{lead-lead collisions, LHC, multiplicity, momentum spectra, interferometry radii}}

%Corresponding author: {\small \textit{Yu.M. Sinyukov, Bogolyubov Institute
%for Theoretical Physics, Kiev 03680, Metrolohichna 14b, Ukraine. E-mail:
%sinyukov@bitp.kiev.ua}}

\section{Introduction}

The realistic simulation of a relativistic heavy ion collision, which allows to describe or predict a wide
variety of measured bulk observables, requires a complicated model consisting of several components, each of
them describing a certain stage of the collision in the most appropriate approach.

According to this requirement, the well known iHKM model~\cite{ihkm,ihkm2} consists of a set of modules/stages, 
corresponding to a series of successive phases of system's evolution process, including formation of superdense matter, its gradual thermalization, 
viscous relativistic hydrodynamic expansion as continuous medium, particlization and the hadronic cascade stage.
The ``development'' of the system at each stage is regulated by specific model parameters, and the aggregated
effect of the matter's evolving during all the stages defines the final state of the system and the behavior of 
various observables, measured in the experiments.

Naturally, a researcher is interested in finding the connection between different parameter values and the
simulation results in order to have the possibility to better understand the nature of 
the investigated processes and to discover the properties and characteristics of hot dense matter, 
basing on the experimental data.

In our previous papers~\cite{ihkm,ihkm2} we have already studied the influence of different iHKM parameters
(such as initial energy-density profile shape and its momentum anisotropy, shear viscosity at the hydrodynamical 
stage etc.) on the obtained results for spectra, $v_2$ coefficients, interferometry radii, etc., 
in application to Pb+Pb collisions at the LHC energy $\sqrt{s_{NN}}=2.76$~TeV. 

In the present paper we are aiming to investigate the dependence of our simulation results on such parameters
as thermalization and relaxation times at the pre-thermal stage, $\tau_{th}$ and $\tau_{rel}$, at the LHC energies, 
and the particlization/hadronization temperatures $T_p$, associated with the different equations of state, 
giving ourselves a task to predict/describe a set of bulk observables in the LHC Pb+Pb collisions 
at the higher energy $\sqrt{s_{NN}}=5.02$~TeV.

\section{Relaxation rate and thermalization time in the LHC collisions within iHKM}

The expansion of thermalized continuous medium near local thermal and chemical equilibrium is described
in iHKM on the basis of the relativistic viscous hydrodynamics approximation in the Israel-Stewart formalism.
This stage of the system's evolution is characterized by the shear viscosity parameter, $\eta/s$.
As it was found in~\cite{ihkm}, this parameter for quark-gluon matter should have the value close
to the minimal possible one, $1/4\pi \approx 0.08$.
The initial conditions for subsequent hydrodynamical evolution of the system are formed during the pre-thermal 
relaxation stage, which models the transformation of initially non-thermalized system into thermalized one \cite{ihkm2}.

The initial pre-thermal states are constructed with the help of MC Glauber GLISSANDO code~\cite{gliss} and are regulated 
by the following model parameters: $\epsilon_0(\tau_0)$ --- the initial maximal energy density
in the center of the system at the starting proper time $\tau_0$, $\alpha$ --- the parameter defining the
proportion between the contributions from ``binary collisions'' and ``wounded nucleons'' models into the
initial energy-density profile, and $\Lambda$ --- regulating the momentum anisotropy of the initial parton
distribution function. 
The values of these parameters are defined from the experimental data on final particle multiplicities and pion spectra. 
It turns out that $\alpha$ and $\Lambda$ at the LHC remain the same at different energies of $Pb+Pb$ collisions and various sets 
of other model parameters. 
The value of $\Lambda$ corresponds to a very large momentum anisotropy of the initial state, that is typical for the models based on Color Glass Condensate (CGC) approach.   

The time parameters, determining the dynamics of the pre-thermal stage, are the initial time $\tau_0$, relaxation time $\tau_{rel}$ and 
thermalization time $\tau_{th}$. The first one defines the proper time, when the initial energy-density profile of strongly interacting matter 
is formed, the second one is related to the rate of transformation of the system
from the initial non-equilibrium state to the hydrodynamical one, and the third parameter defines the time when the 
system becomes nearly thermal.

After the pre-thermal and hydrodynamical stages, the particlization stage follows, since the continuous medium description 
becomes inappropriate, and the system must be considered as the set of hadrons. This stage is characterized by the particlization 
temperature $T_p$ (close to the temperature of hadronization), from which it starts. 
The temperature $T_p$ is determined by the QCD equation of state and is defined as the temperature
when the quark-gluon matter mostly transforms into the hadron-resonance gas.

At the last, hadronic cascade stage, the particles created during the particlization, collide with each other,
experiencing elastic and inelastic scatterings, and also the resonance decays take place. This stage is simulated
within UrQMD model~\cite{urqmd}. Here one can
allow or forbid certain decays (e.~g., in order to reproduce the experimental feed-down treatment etc.),
switch off some processes, like baryon-antibaryon annihilation and so on.

In the article~\cite{ihkm} it was determined that the behavior of bulk observables in iHKM depends strongly on
the time of the initial state formation, $\tau_0$. 
As for the relaxation time $\tau_{rel}$ and thermalization time $\tau_{th}$ characterizing the intensity
of the thermalization process, this issue remained investigated insufficiently. 
To clarify this dependence in the current study we compare the iHKM results on the particle 
$p_T$ spectra at different $\tau_{rel}$ and $\tau_{th}$, having fixed the initial
time $\tau_0=0.1$~fm/$c$ and simultaneously changing the maximal initial energy density $\epsilon_0(\tau_0)$. 

\begin{figure}
\centering
\includegraphics[bb=0 0 567 409, width=12.5cm]{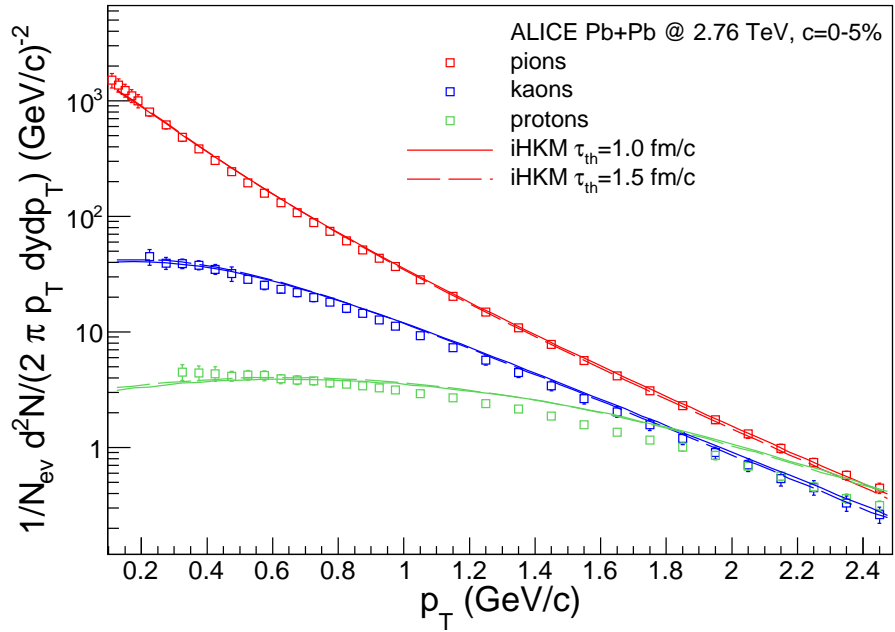}
\caption{The iHKM results on pion, kaon and proton spectra in the Pb+Pb collisions with $c=0-5\%$ at the LHC 
energy $\sqrt{s_{NN}}=2.76$~TeV. The two different thermalization time values, $\tau_{th}=1.0$~fm/$c$ (solid lines) and $\tau_{th}=1.5$~fm/$c$ (dashed lines) were utilized. The experimental data from the ALICE Collaboration~\cite{alice} are shown as square markers.
\label{taus1}} 
\end{figure}

First of all, we fix the relaxation time $\tau_{rel}$=0.25~fm/$c$ and compare $p_T$ spectra for main
particle species, obtained in the model at the two thermalization times $\tau_{th}=1.0$ fm/$c$ and 
$\tau_{th}=1.5$ fm/$c$ for Pb+Pb collisions at the LHC energy $\sqrt{s_{NN}}=2.76$~TeV. 
To ensure the right reproduction of the mean charged particle
multiplicity in the model for all centralities in both cases, we use two different $\epsilon_0$ values: 
$\epsilon_0=834$~GeV/fm$^3$ for $\tau_{th}=1.0$ fm/$c$ and $\epsilon_0=681$~GeV/fm$^3$ for $\tau_{th}=1.5$ fm/$c$. 
The corresponding results almost coincide: the comparison can be seen in Fig.~\ref{taus1}, where the iHKM pion, kaon and proton transverse momentum spectra are presented for the case of $c=0-5\%$ events together with the ALICE Collaboration~\cite{alice}
experimental data.

\begin{figure}
\centering
\includegraphics[bb=0 0 567 409, width=12.5cm]{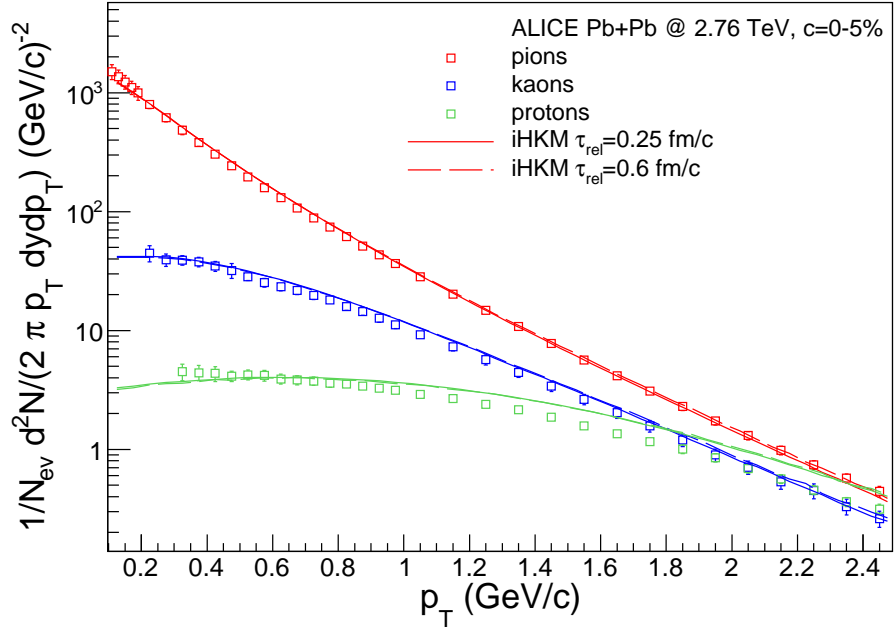} 
\caption{The comparison of the iHKM results on pion, kaon and proton spectra in the Pb+Pb collisions with 
$c=0-5\%$ at the LHC energy $\sqrt{s_{NN}}=2.76$~TeV. The two different relaxation times, $\tau_{rel}=0.25$~fm/$c$ (solid lines) and $\tau_{rel}=0.60$~fm/$c$ (dashed lines) were used. 
The thermalization time value in both cases was set to $\tau_{th}=1.5$~fm/$c$.
The experimental data measured by the ALICE Collaboration~\cite{alice} are shown as square markers.
\label{taus2}} 
\end{figure}

In the next step we, on the contrary, fix the value of the thermalization time at $\tau_{th}=1.5$~fm/$c$,  
and calculate the spectra at the two values of the relaxation time, $\tau_{rel}=0.25$ fm/$c$ and 
$\tau_{rel}=0.6$~fm/$c$ (here we choose the second $\tau_{rel}$ value to be noticeably larger than the first one,
but  smaller than the time of thermalization). Again we re-tune also the value of the maximal
initial energy density, setting it to 630~GeV/fm$^3$ for the case of $\tau_{rel}=0.6$~fm/$c$.
The Fig.~\ref{taus2} shows the compared results for the two $\tau_{rel}$, again together with the experimental 
points.

Both comparisons of the results in Figs.~\ref{taus1}, \ref{taus2} demonstrate that while the maximal initial
energy density $\epsilon_0$ remains a free parameter, the experimentally measured $p_T$ spectra
can be successfully described at different thermalization and relaxation times, characterizing
the rate of the matter's thermalization process. This fact complicates the experimental study of
the process of thermalization in heavy ion collisions.

\section{Results for $\sqrt{s_{NN}}=5.02$ TeV and discussion}

At first in order to adjust the model parameters to the description of Pb+Pb collisions at the LHC energy 
$\sqrt{s_{NN}}=5.02$~TeV we fitted the mean charged particle multiplicity dependence on centrality and the slope of the pion 
$p_T$ spectrum at the two 
chosen particlization temperatures and corresponding equations of state for quark-gluon phase.
We considered two cases: the particlization temperature $T_p=165$~MeV with the Laine-Schroeder equation of state~\cite{EoS} and $T_p=156$~MeV with the HotQCD Collaboration EoS~\cite{EoS2}.
For $T_p=165$~MeV the best fit to multiplicity dependence corresponds to $\epsilon_0=1067$~GeV/fm$^3$ at the 
initial time for the system formation $\tau_0 = 0.1$~fm/$c$, and for $T_p=156$~MeV the values $\epsilon_0=870$~GeV/fm$^3$ and $\tau_0 = 0.12$~fm/$c$ are obtained from the fit. 
The different initial times $\tau_0$ were used to optimize the description of pion transverse momentum 
spectrum at low $p_T$.
The parameter regulating the proportion between wounded nucleons and binary collisions 
model contributions to the initial energy density profile from GLISSANDO code~\cite{gliss} for 
both particlization temperatures is $\alpha=0.24$ --- the same as for the energy $\sqrt{s_{NN}}=2.76$~TeV. 

The rest of the model parameters 
--- the thermalization time $\tau_{th} = 1$~fm/$c$, the relaxation time at the pre-thermal stage $\tau_{rel} = 0.25$~fm/$c$, and the momentum anisotropy of the initial 
state $\Lambda=100$ --- also have the values equal to those provided a successful description of Pb+Pb collision 
observables at $\sqrt{s_{NN}}=2.76$~TeV~\cite{ihkm,ihkm2}. 

\begin{figure}
\centering
\includegraphics[bb=0 0 567 409,width=0.88\textwidth]{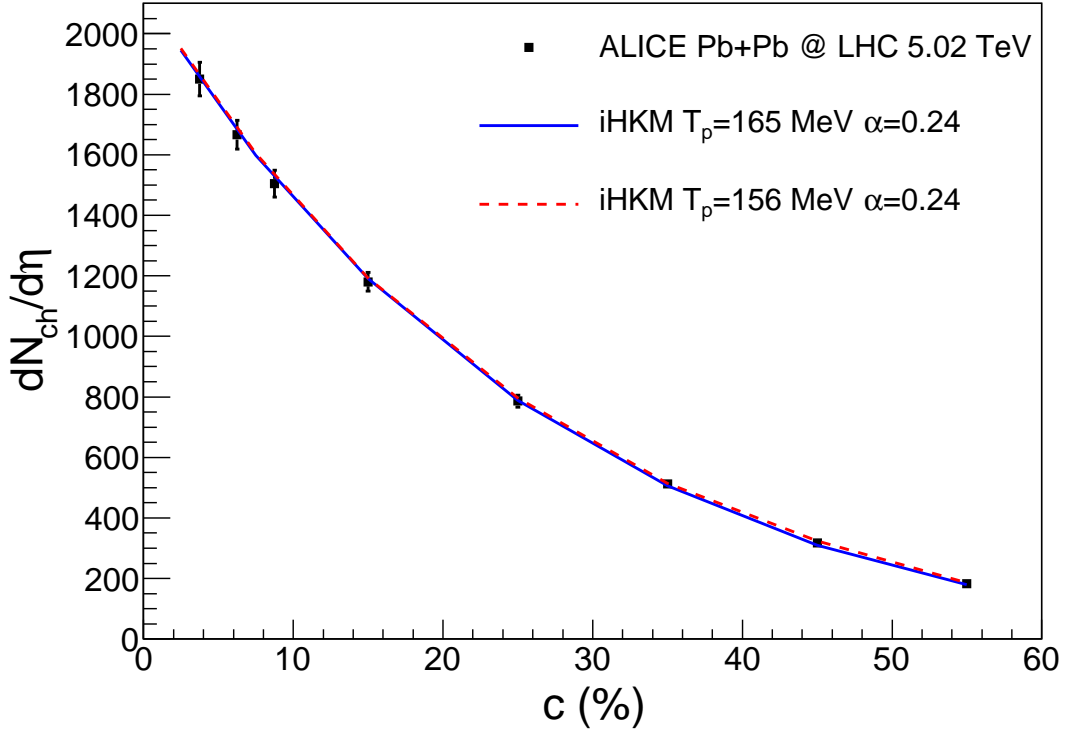}
  \caption{The mean charged particle multiplicity $dN_{ch}/d\eta$ dependence on collision centrality
  for Pb+Pb collisions at the LHC energy $\sqrt{s_{NN}}=5.02$~TeV. 
  The best iHKM descriptions at $T_p=165$~MeV and $T_p=156$~MeV of the ALICE experimental data~\cite{mult} are used to fix the main model parameters $\epsilon_0$ and $\alpha$ in both cases. 
\label{mult}}
\end{figure}

The multiplicity descriptions obtained at chosen parameter values for $\sqrt{s_{NN}}=5.02$~TeV collisions 
in comparison to the experimental data are presented in Fig.~\ref{mult}.

\begin{figure}
\centering
\includegraphics[bb=0 0 567 409, width=0.8\textwidth]{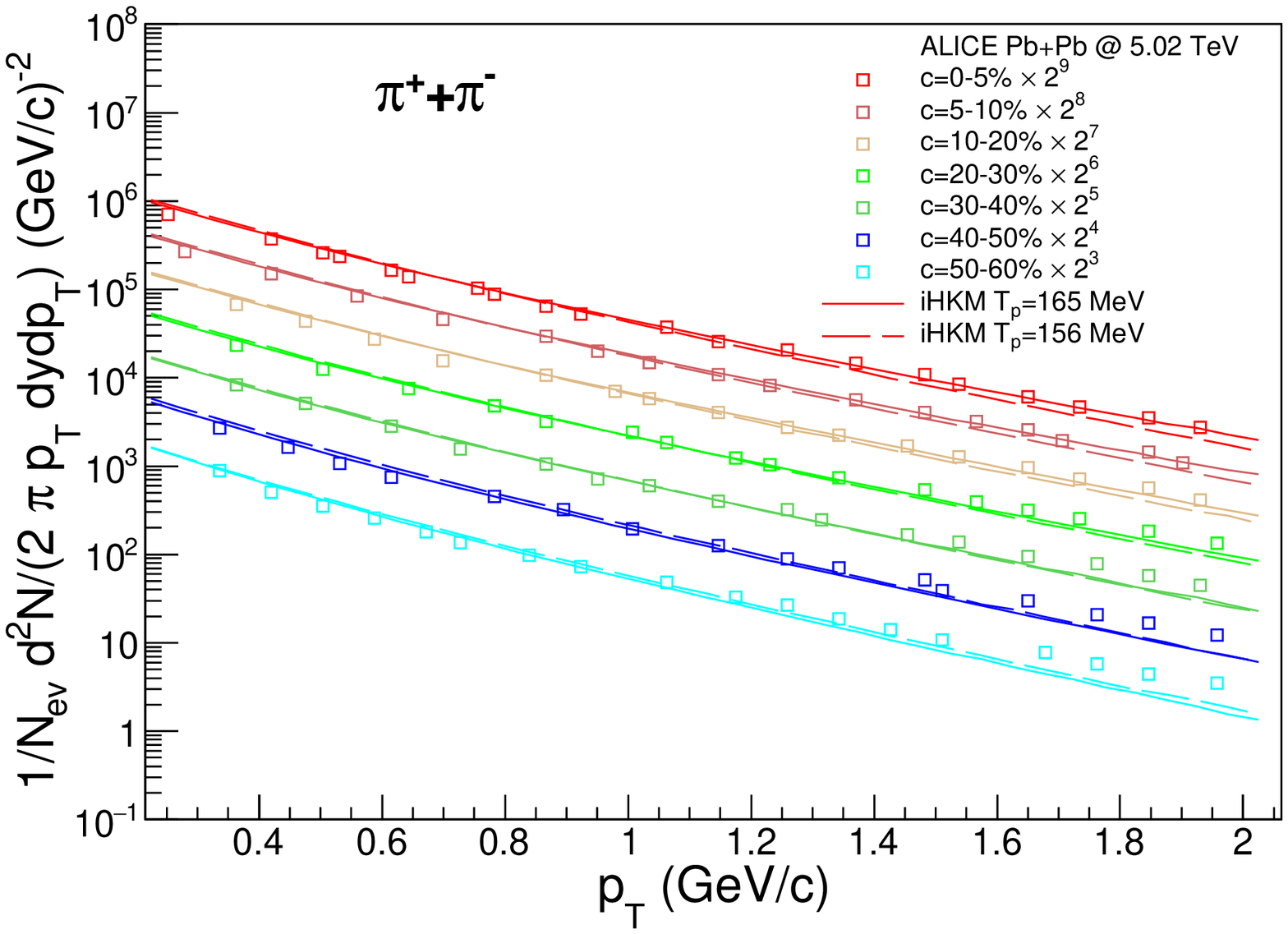} \\
\includegraphics[bb=0 0 567 409, width=0.8\textwidth]{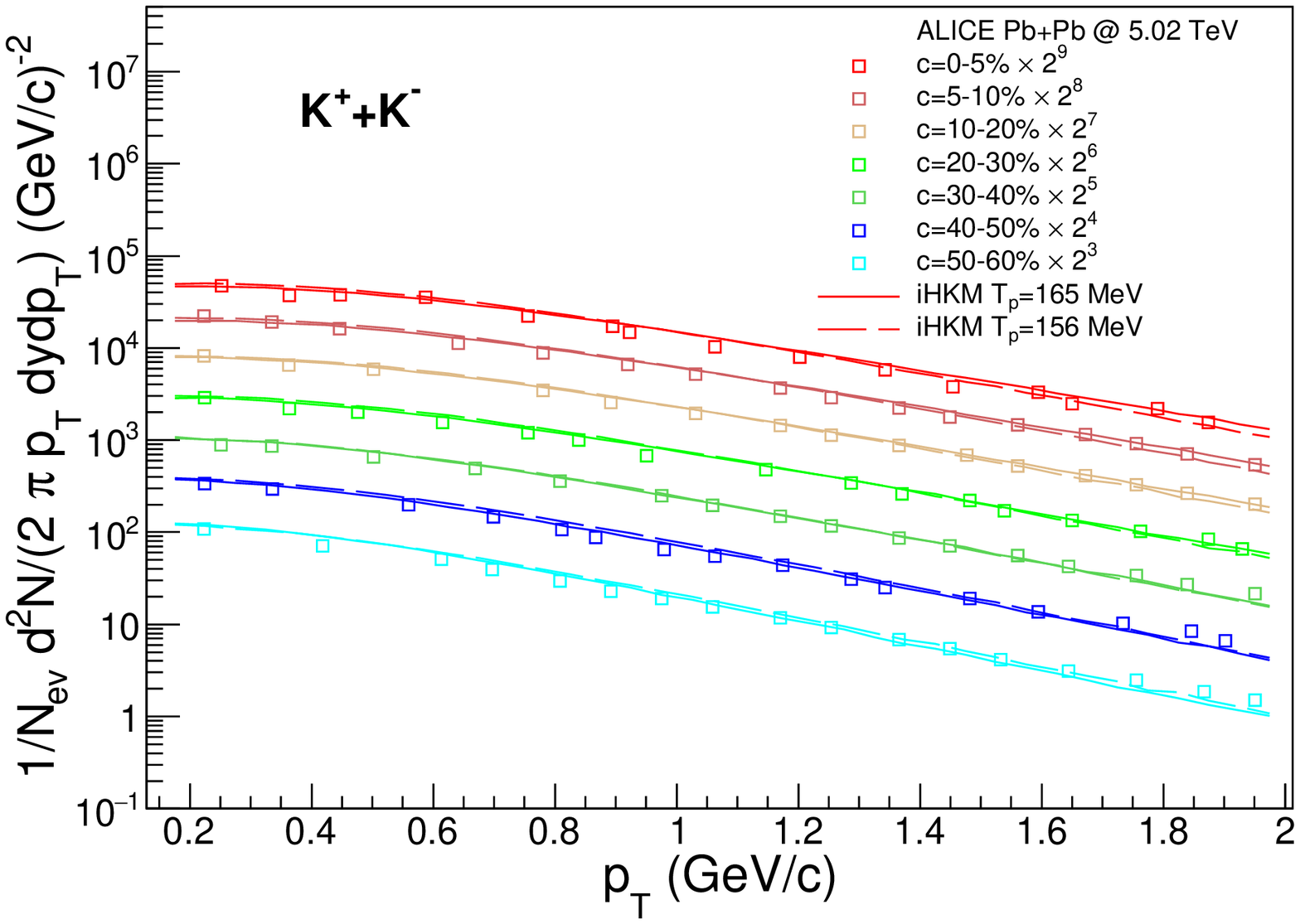} 
\caption{Pion and kaon transverse momentum spectra for different centrality
classes obtained from iHKM at the two particlization temperatures in comparison with the preliminary
experimental data from the ALICE Collaboration~\cite{spec502}. For better visibility the spectra points 
for different centralities are scaled by different powers of 2.
\label{specs}} 
\end{figure}

\begin{figure}
\centering
\includegraphics[bb=0 0 567 409, width=0.8\textwidth]{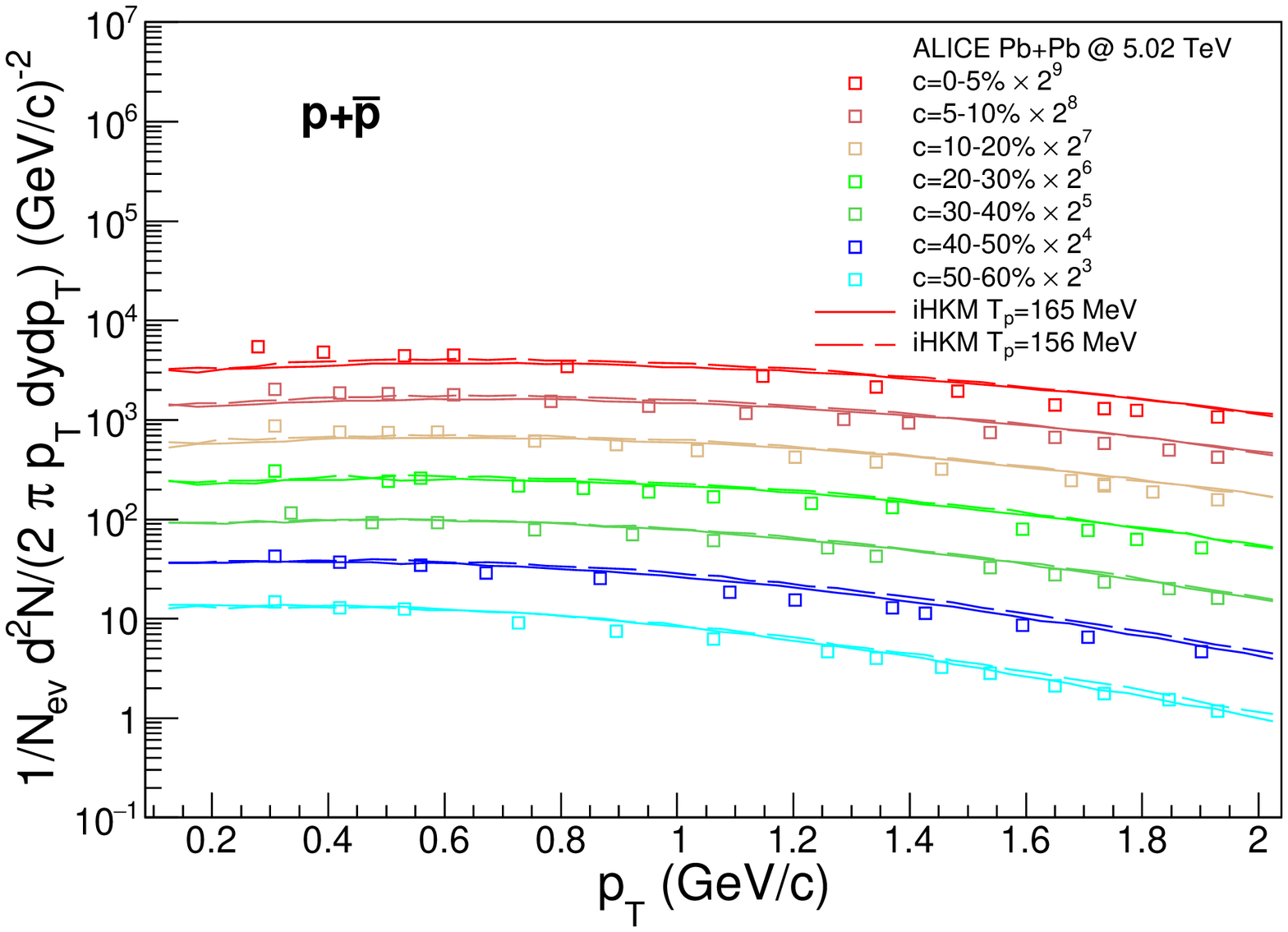}
\caption{The same as in Fig.~\ref{specs} for protons.
\label{specs2}}
\end{figure}

Using the found values of model parameters, we calculated also the transverse momentum spectra of pions, kaons and 
protons for different centralities (see Figs.~\ref{specs}, \ref{specs2}). The model simulation results are 
compared to the preliminary experimental data on $p_T$ spectra measurement, presented at Quark Matter 2017~\cite{spec502}.
As one can see, iHKM gives a satisfactory description of the particle production for the $\sqrt{s_{NN}}=5.02$~TeV
Pb+Pb collisions in both $T_p/$EoS modes. 

It is worth noting that the pion spectra in iHKM model is described within the experimental errors in a wide $p_T$ region including soft momentum 
interval. Previously the similar result was observed and emphasized in Ref.~\cite{ihkm2} for the LHC energy $\sqrt{s_{NN}}=2.76$~TeV. Therefore, in iHKM 
there is no necessity to include a specific mechanism for soft pion radiation, for example, through the Bose-Einstein condensation~\cite{Begun}, 
in order to describe the soft pion emission in Pb+Pb collisions at the LHC.

\begin{figure}
\centering
\includegraphics[bb=0 0 567 409, width=0.9\textwidth]{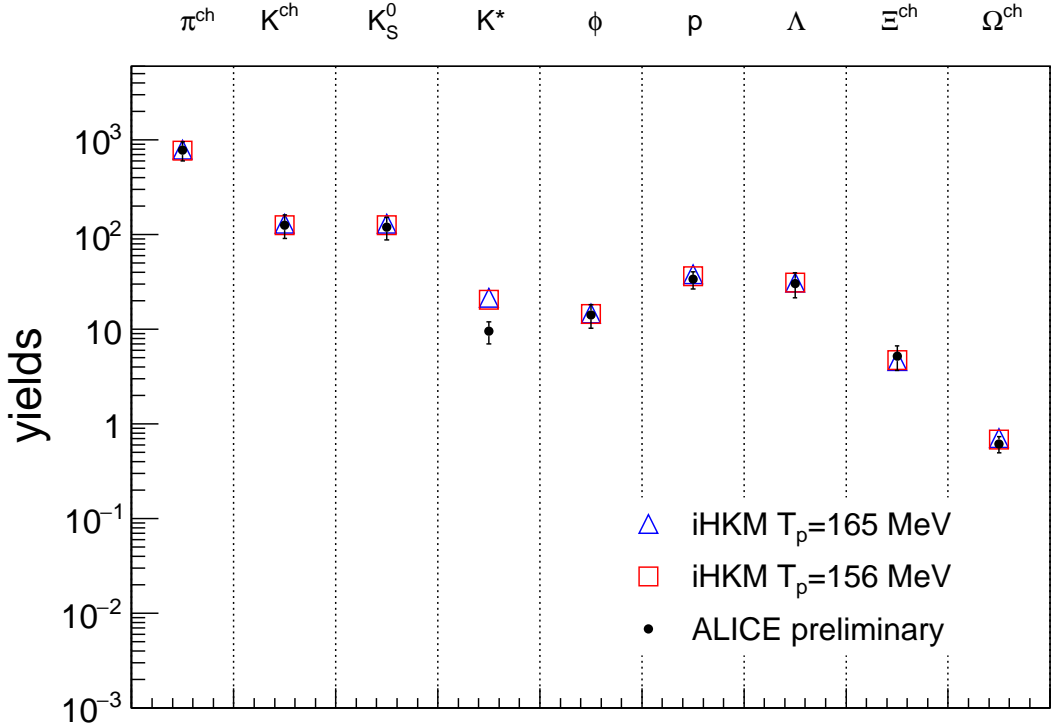}
\caption{Particle yields $dN/dy$ in events with $c=0-10\%$ calculated in iHKM compared to the preliminary ALICE data~\cite{enrico1,enrico2}.
\label{yields}}
\end{figure}

In Fig.~\ref{yields} one can see the $p_T$-integrated particle yields $dN/dy$ for different species,
calculated in the model in comparison with the preliminary experimental data, provided by the ALICE Collaboration
and reported in talks at Quark Matter 2018 and WPCF 2018~\cite{enrico1,enrico2}.
 

\begin{figure}
\centering
\includegraphics[bb=0 0 567 409, width=0.9\textwidth]{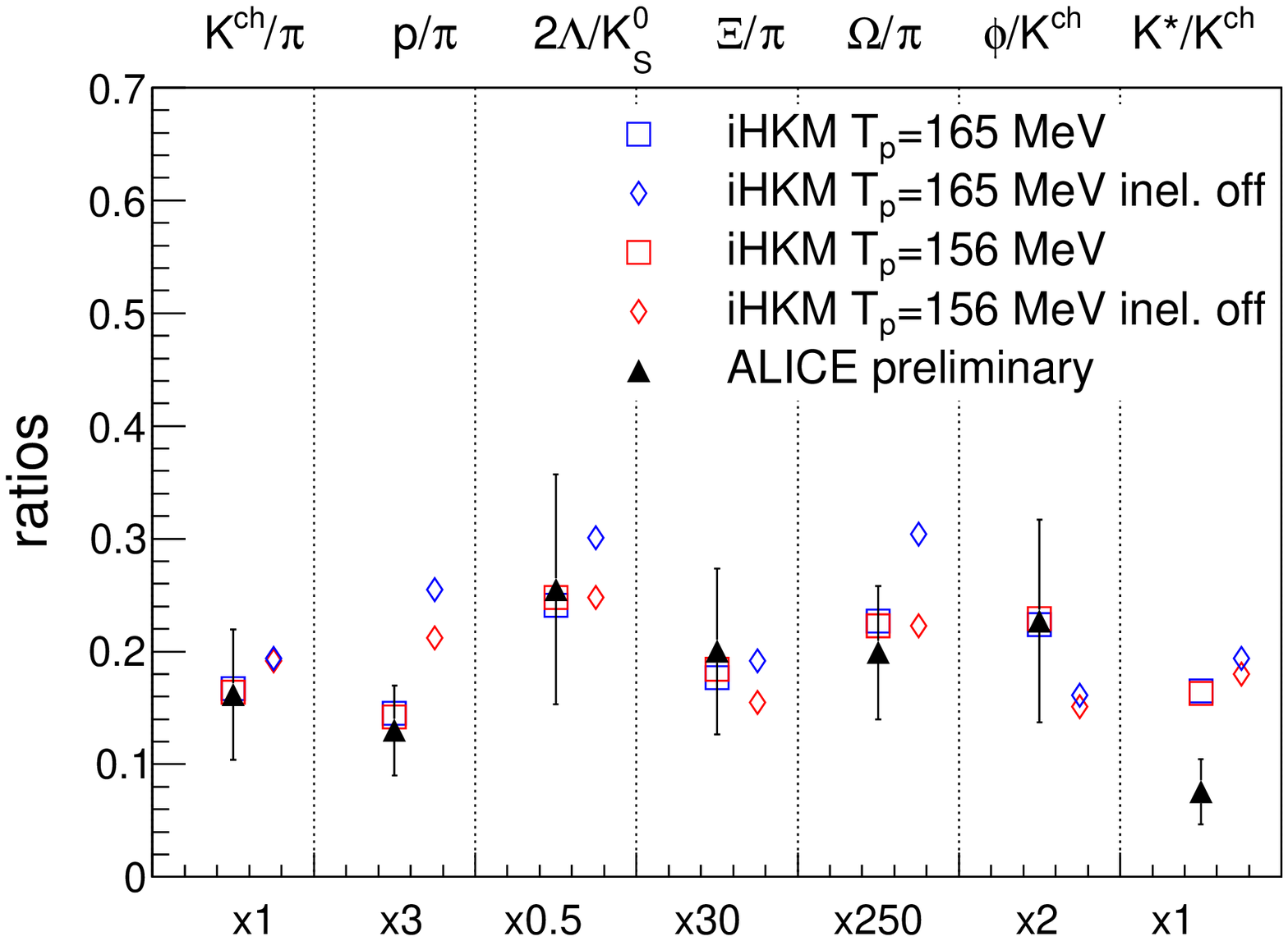}
\caption{The iHKM results for the particle number ratios calculated at the two particlization temperatures
$T_p=165$~MeV and $T_p=156$~MeV for collisions with $c=0-10\%$ compared to the preliminary ALICE data~\cite{enrico1,enrico2}. The simulations are performed in two regimes: the full one and the one with inelastic processes turned off. 
 \label{ratios}}
\end{figure}

In Fig.~\ref{ratios} we demonstrate the results of iHKM simulations for the various particle number ratios
in central Pb+Pb collisions ($c=0-10\%$) at the considered LHC energy again together with the preliminary
experimental results from the ALICE Collaboration~\cite{enrico1,enrico2}. Here we present the ratios obtained 
from the full iHKM calculations along with those calculated in the mode with inelastic reactions turned off.
As one can see, while the results for certain ratios in the latter ``reduced'' mode can noticeably differ for the two particlization 
temperatures, the full calculation always gives very close values in both cases. This peculiarity in
results is in accordance with that observed and discussed in our previous work~\cite{ratiosour}. 
The compensatory mechanism, eliminating in the ``full'' mode the difference, observed between the ratios at the two EoS/$T_p$'s 
in the ``reduced'' mode, is associated with inelastic reactions in the hadronic cascade, which last longer at a higher particlization temperature. 
The model points are in agreement with the preliminary data, except for $K^{*}/K^{ch}$ ratio.

Turning back to the Fig.~\ref{yields}, one sees that the latter discrepancy is due to 
overestimation of the $K^{*}$ yield in the model --- its value is approximately twice higher
in iHKM, than in the experiment. This fact looks strange, since in our recent article~\cite{ratiosour},
focused on the particle production in the LHC Pb+Pb collisions at lower energy, $\sqrt{s_{NN}}=2.76$~TeV,
we obtained a good agreement between model $K^{*}$ yield description and the corresponding ALICE data.
Note, that in our simulations we accounted also for possible problems with the $K^{*}$ identification 
through the products of its decay into kaons and pions, which could be caused by the interaction 
of these daughter particles with the surrounding hadronic medium.
This issue was analyzed in detail in another our paper~\cite{kstar}.
There we considered the case of Pb+Pb collisions at the LHC at $\sqrt{s_{NN}}=2.76$~TeV and 
performed the simulations in iHKM, applying the experimental procedure of $K^{*}$ identification,
i.e. selecting the $K\pi$ pairs, emitted from the close space points and having the specific invariant mass,
corresponding to $K^{*}(892)$ resonance. Then we compared in such a way obtained $K^{*}$ yields with the 
full numbers of $K^{*}$, generated in the model during the hadronization of continuous quark-gluon medium. 
We found that the observed number of $K^{*}$ resonances, reconstructed via the products of their decays, 
for the most central events can be up to 20\% less than the actual number of ``primary'' $K^{*}$'s.
But it is worth noting, that the mentioned reduction is a result of interplay of two opposite processes --- 
the scattering of daughter $K$ and $\pi$, resulting in an impossibility to detect the respective 
parent $K^{*}$'s (this affects up to 70\% of direct resonances), and the recombination of $K\pi$ pairs, producing 
additional $K^{*}$ resonances (this mechanism can give up to 50\% of the initial number of direct $K^{*}$'s).
Both these processes in iHKM are simulated at its final ``afterburner'' stage within the UrQMD model.

In the present study we apply the same $K^{*}$ restoration procedure, but this time obtain a higher
$dN/dy$, than in the data (note also that this preliminary experimental $dN/dy$ value for $K^{*}$ at $\sqrt{s_{NN}}=5.02$~TeV
is noticeably smaller than that at $\sqrt{s_{NN}}=2.76$~TeV, while the iHKM results at both collision energies are close).
The reason for a low value of $K^{*}$ yield in the preliminary experimental results is unclear and 
needs further investigation. It can be possibly explained by the identification problems in the
experiment and one cannot exclude that the $dN/dy$ value for $K^{*}$ can be corrected in the final version of ALICE data.

\begin{figure}
\centering
\includegraphics[bb=0 0 567 409, width=0.49\textwidth]{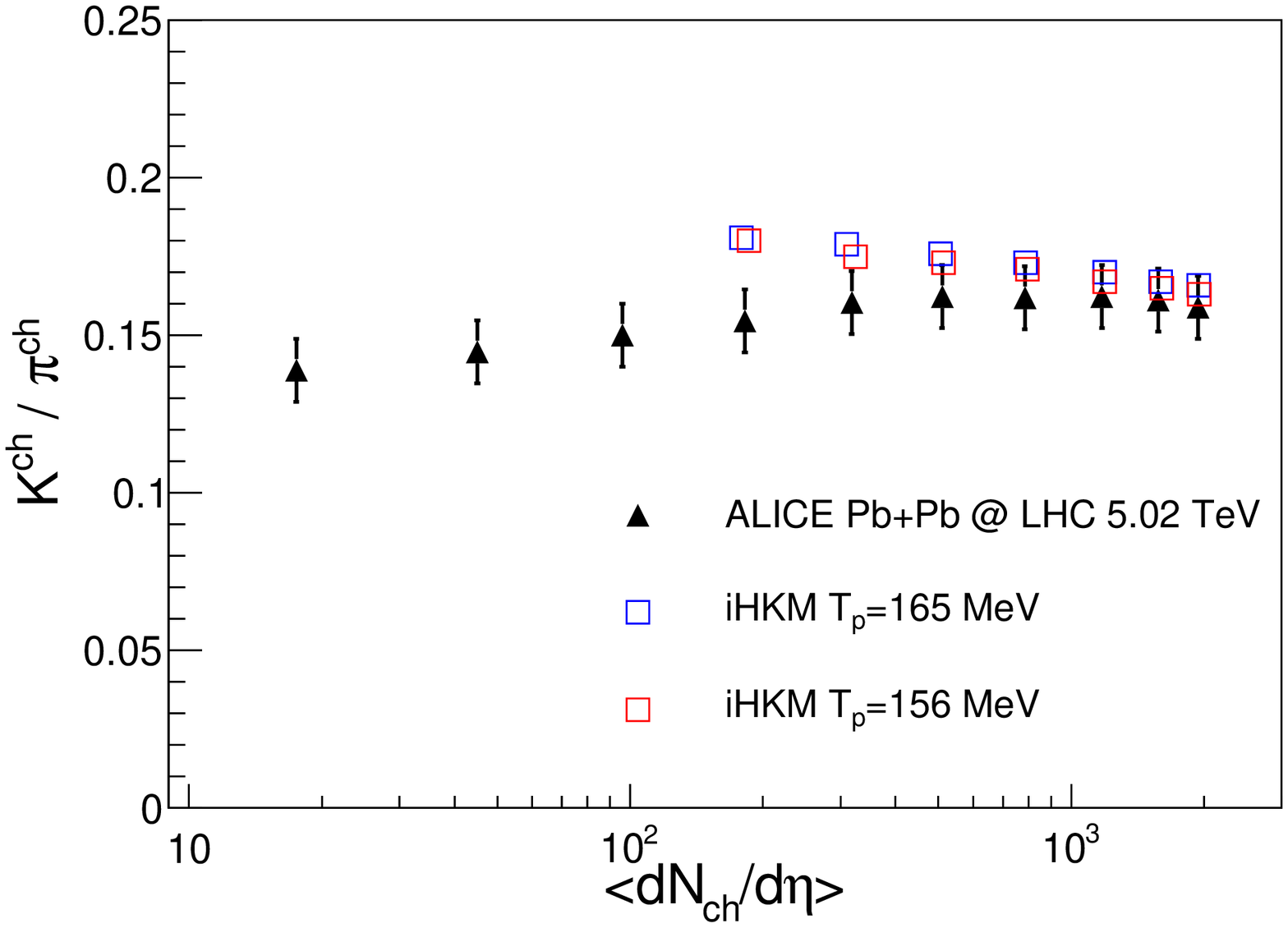}
\includegraphics[bb=0 0 567 409, width=0.49\textwidth]{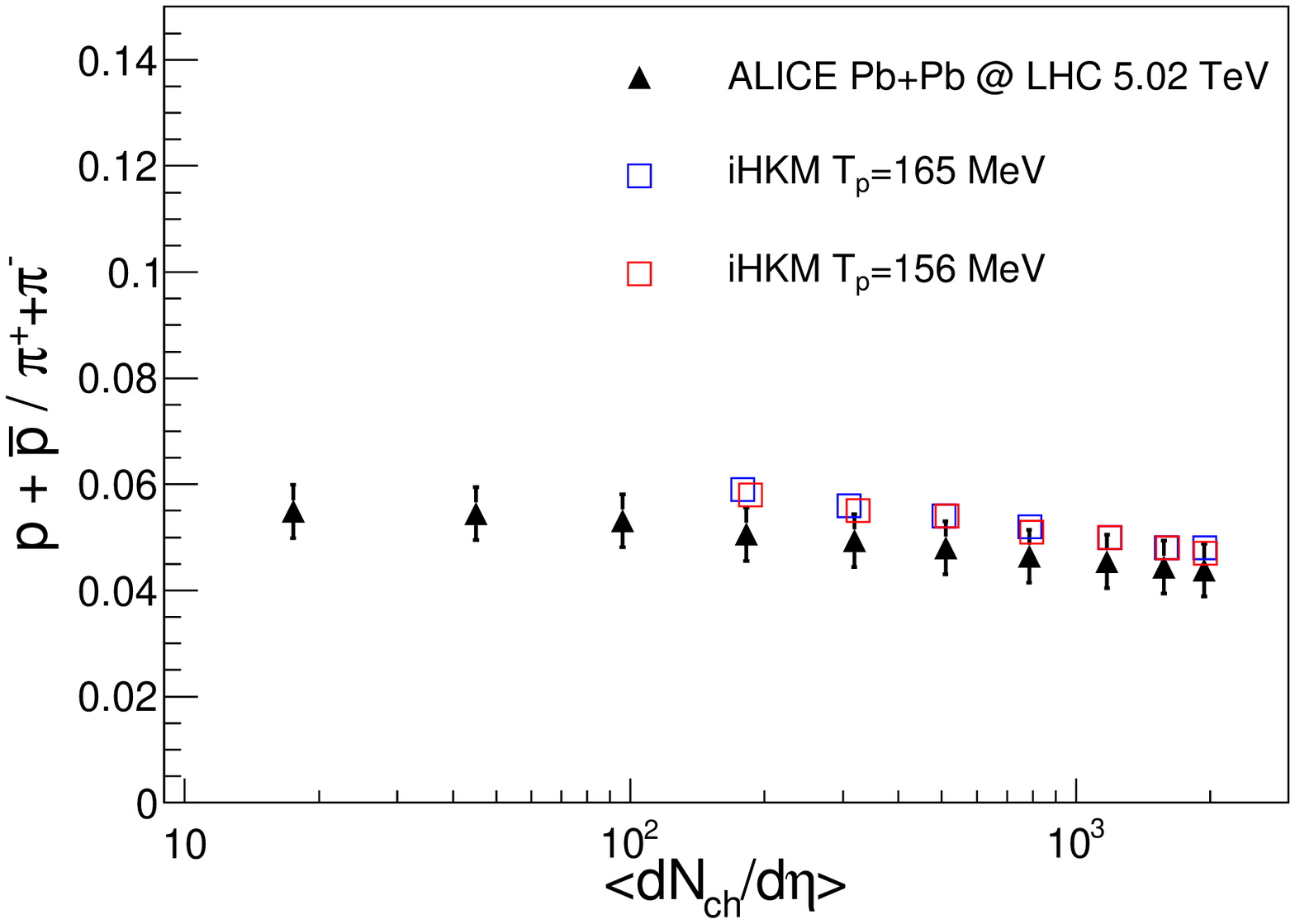} 
\caption{The iHKM results for the $K/\pi$ and $p/\pi$ particle number ratios dependence on mean charged particle multiplicity calculated at particlization temperatures $T_p=165$~MeV and $T_p=156$~MeV and the corresponding experimental data~\cite{spec502}. 
\label{ratiosmult}} 
\end{figure}

\begin{figure}
\centering
\includegraphics[bb=0 0 567 409, width=0.8\textwidth]{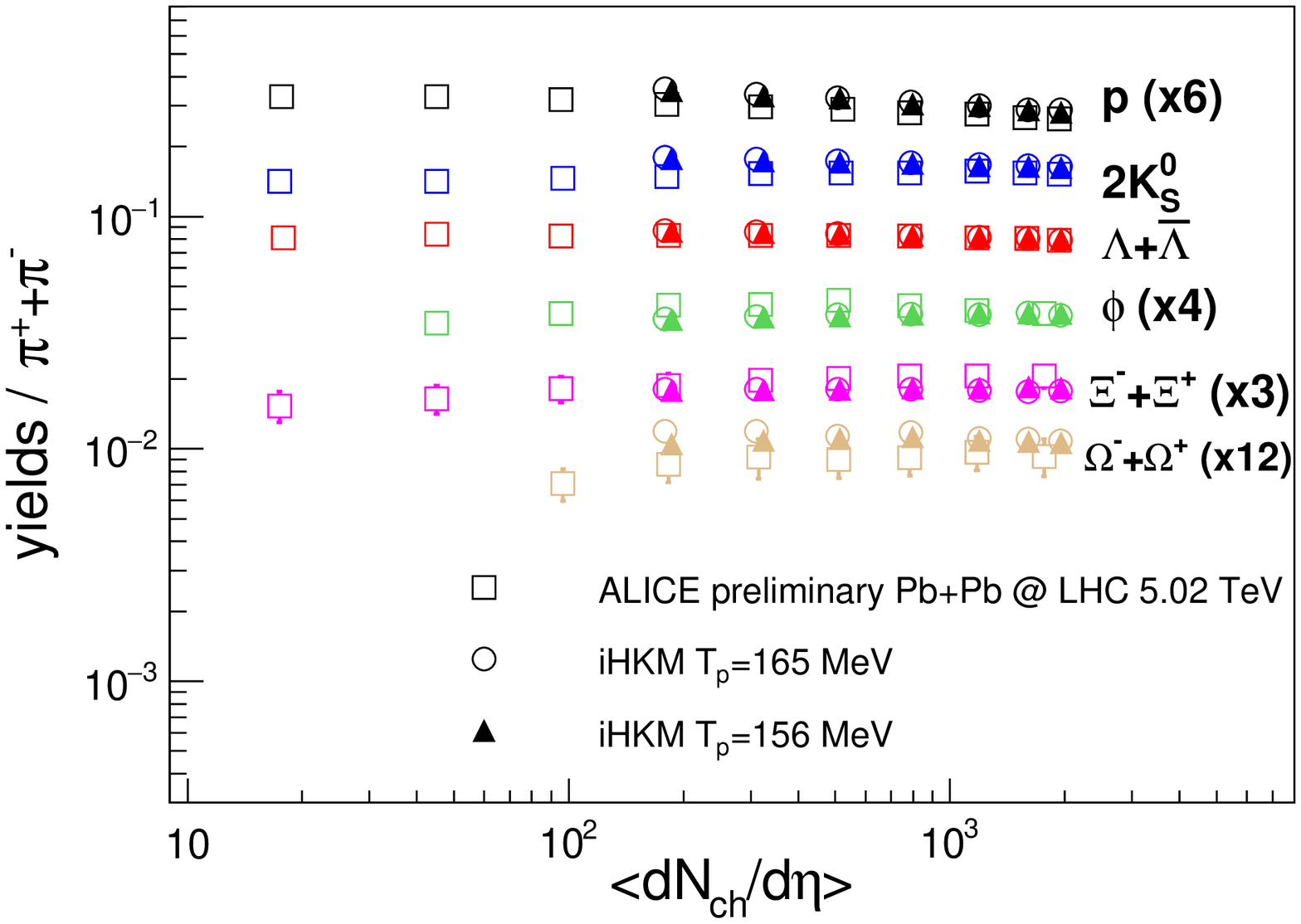}
\caption{The iHKM results together with the preliminary ALICE Collaboration data~\cite{enrico2} on the behavior of the ratios of 
different particle yields to the pion yield with centrality, for the LHC Pb+Pb collisions at $\sqrt{s_{NN}}=5.02$~TeV.
 \label{ratiosmult2}}
\end{figure}

For the $K/\pi$ and $p/\pi$ ratios we have also studied within iHKM their dependence on the mean charged particle 
multiplicity for different centrality classes and compared the simulation results to the preliminary experimental 
data, presented in~\cite{spec502} (see Fig.~\ref{ratiosmult}). In the plot one can see that the model describes  
well (in the ``full regime'') both ratios' behavior for not far from central events at both EoS/$T_p$.
The theoretical values obtained in the ``reduced regime'' without inelastic scatterings are fairly larger than
the measured ones. The similar situation can be observed in Fig.~\ref{ratiosmult2}, where the model calculations
of the ratios of various particle species yields to the pion yield are compared to the preliminary experimental
data from the ALICE Collaboration~\cite{enrico2}.

In the next figures we demonstrate our predictions as for the pion and kaon interferometry radii $R_{out}$,
$R_{side}$ and $R_{long}$ measured at different mean pair transverse momentum $k_T$.
The predictions are provided for the three centrality classes: $c=0-5\%$, $c=20-30\%$, and $c=40-50\%$.
The radii behavior demonstrates the approximate $k_T$ scaling at $k_T>0.4$~GeV/$c$, repeating the behavior
observed for the collisions at $\sqrt{s_{NN}}=2.76$~TeV~\cite{Ourkt}. 
The radii's absolute values decrease when going from central to non-central events.

\begin{figure}
\centering
\includegraphics[bb=0 0 567 409, width=0.85\textwidth]{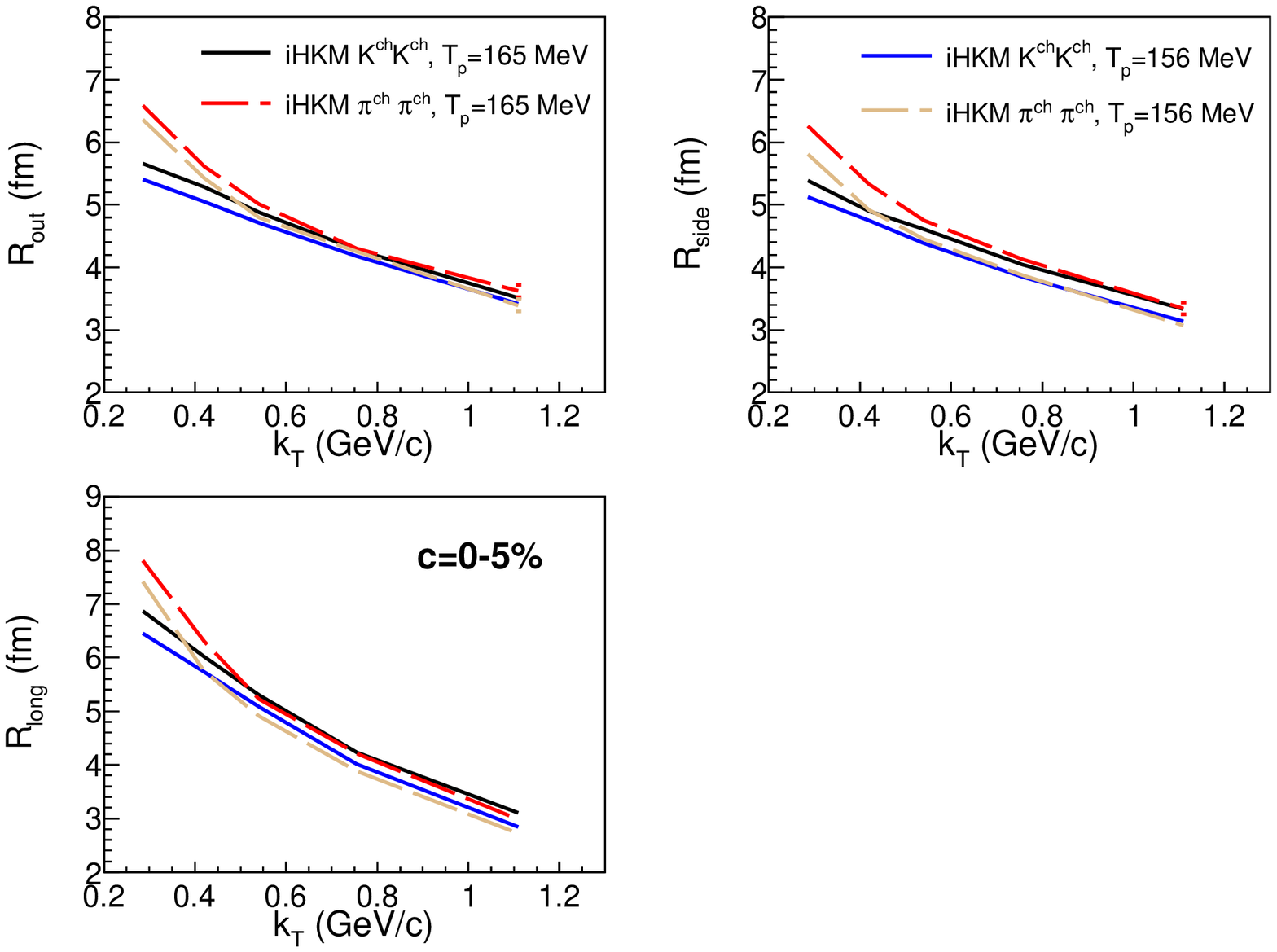}
\caption{The iHKM prediction of the charged pion and kaon interferometry radii $k_T$ dependence for the centrality $c=0-5\%$. The calculations were performed at the two particlization temperatures, $T_p=165$~MeV and $T_p=156$~MeV.
 \label{rad05}}
\end{figure}

\begin{figure}
\centering
\includegraphics[bb=0 0 567 409, width=0.85\textwidth]{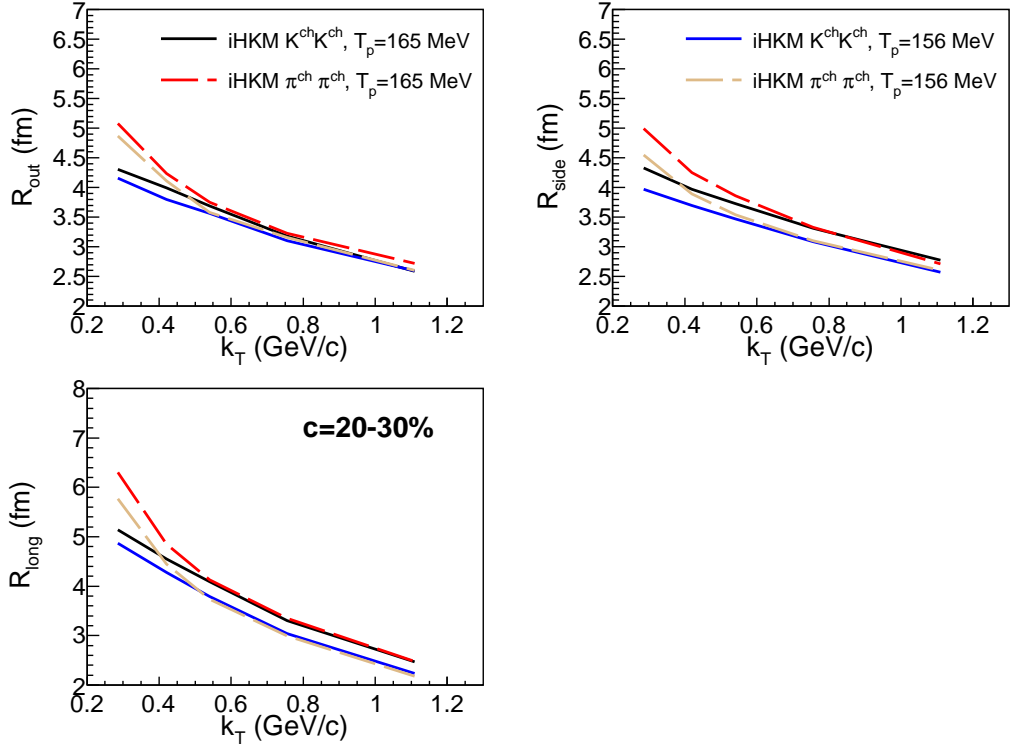}
\caption{The same as in Fig.~\ref{rad05} for the centrality $c=20-30\%$.
 \label{rad2030}}
\end{figure}

\begin{figure}
\centering
\includegraphics[bb=0 0 567 409, width=0.85\textwidth]{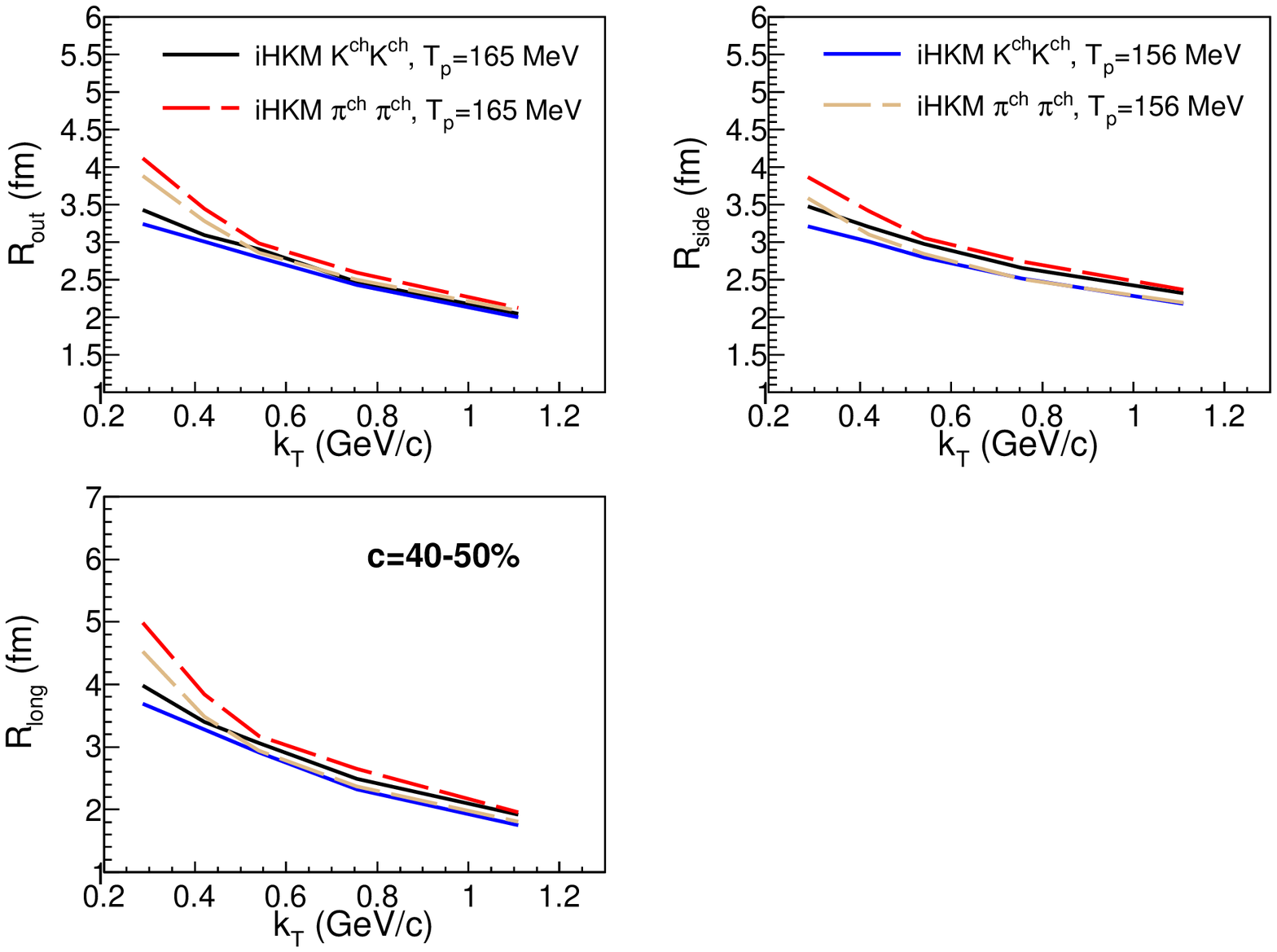}
\caption{The same as in previous two figures for the centrality $c=40-50\%$.
 \label{rad4050}}
\end{figure}

\begin{figure}
\centering
\includegraphics[bb=0 0 567 409, width=0.85\textwidth]{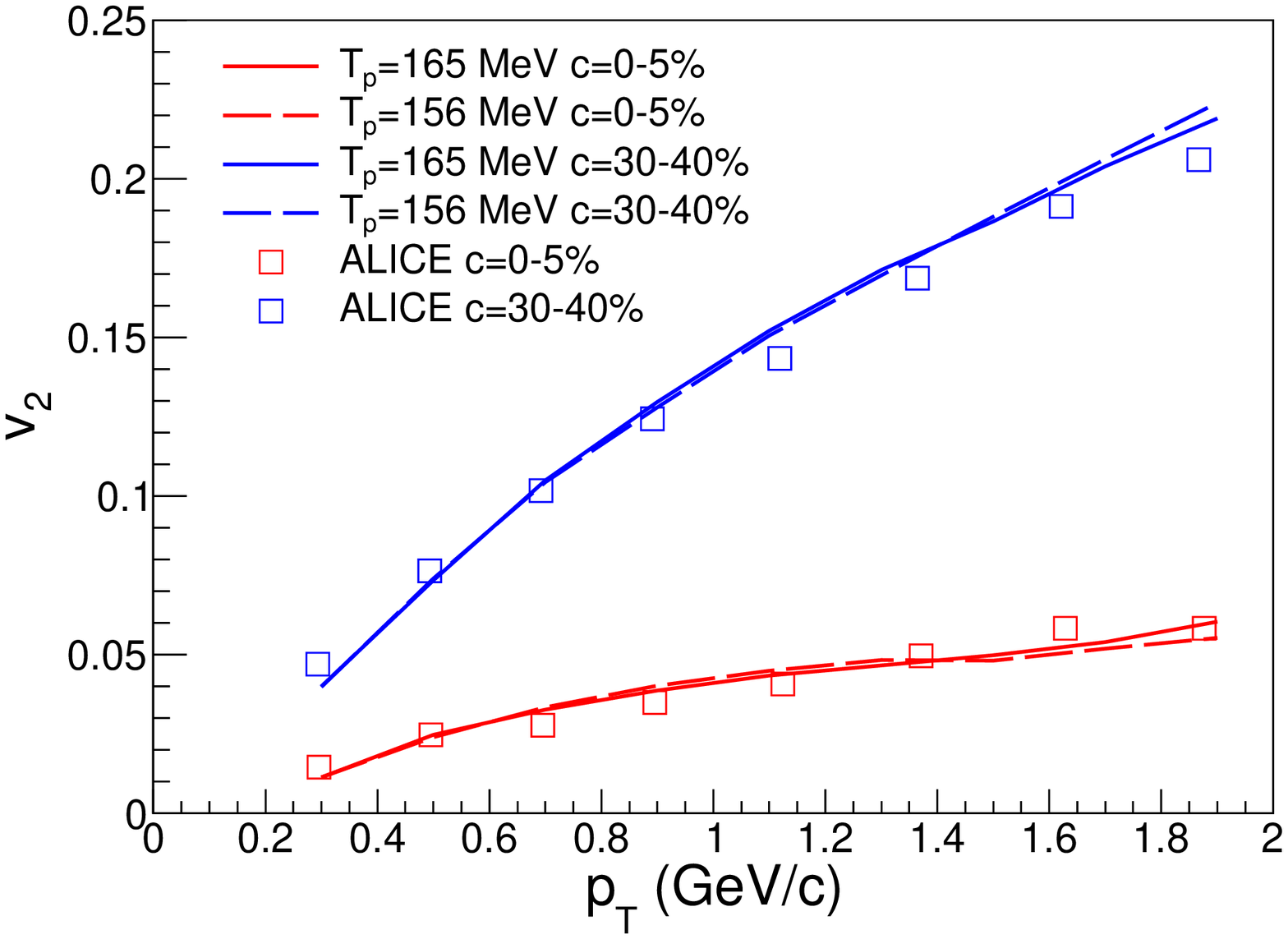}
\caption{The iHKM results on the elliptic flow $v_2$ dependence on $p_T$ for all charged particles together with the corresponding 
ALICE data~\cite{v2alice} for the centrality classes $c=0-5\%$ and $c=30-40\%$. 
The model curves for the two particlization temperatures, $T_p=165$~MeV and $T_p=156$~MeV, are presented. 
 \label{v2}}
\end{figure}

In Fig.~\ref{v2} one can see the comparison of iHKM results on the all charged particles 
elliptic flow $v_2(p_T)$ behavior for the two centrality classes with the ALICE experimental data~\cite{v2alice}.

As one can see, the iHKM results on all the considered bulk observables at the two particlization temperatures
are close to each other, which means that constructing the model for a high-energy collision process, one has 
a certain freedom in choosing the equation of state, particlization temperature and maximal initial energy density, since different combinations of these parameter choices give similar results. This also means that 
experimentally one is hardly able to strictly define each of these parameters separately basing on the considered 
observables. 

\section{Conclusions}

The theoretical predictions and the description of the preliminary experimental results for bulk observables
in Pb+Pb collisions at the LHC energy $\sqrt{s_{NN}}=5.02$~TeV within iHKM model are presented.
The results' dependence on the model parameters is investigated. In particular, the model simulations were
performed using two different particlization temperatures with two corresponding equations of state for 
quark-gluon phase. The comparison of the obtained results in these two cases shows that both $T_p$/EoS can provide
equally good description of measured data and quite close predictions for not yet measured observables,
if having changed the $T_p$, one simultaneously re-adjusts the maximal initial energy density parameter 
$\epsilon_0(\tau_0)$. This result confirms our previous observation \cite{ratiosour}  as for particle number ratios for
Pb+Pb collisions at lower LHC energy $\sqrt{s_{NN}}=2.76$~TeV.

As for the effect of changing such characteristics of the pre-thermal stage as the thermalization time $\tau_{th}$ and relaxation time $\tau_{rel}$ on simulation
results, we demonstrate in this article that varying these parameters in quite wide ranges ($1-1.5)$~fm/$c$ for 
$\tau_{th}$ and $(0.25-0.6)$~fm/$c$ for $\tau_{rel}$ accompanied again by the corresponding change of $\epsilon_0(\tau_0)$
does not influence the final particle $p_T$ spectra. Thus, one can conclude that some important characteristics of the
matter evolution in $A+A$ collisions cannot be definitely extracted from the experimental results, if the maximal initial energy 
density (or --- in other than iHKM kinds of models --- a combination of some other parameters that set the model) is undefined 
and so can be considered as (effective) free parameter.

%\section*{Acknowledgments}
\begin{acknowledgments}
The authors thank P. Braun-Munzinger for initializing this article, his careful reading of the manuscript and valuable comments.
The research was carried out within the scope of the European Research Network ``Heavy ions at ultrarelativistic energies'' and corresponding Agreement 
with the National Academy of Sciences (NAS) of Ukraine. The work is partially supported by 
the NAS of Ukraine Targeted research program ``Fundamental research on high-energy physics and nuclear physics (international cooperation)'', Agreement 7-2018.  
\end{acknowledgments}

\end{document}